\documentclass[aps,pra,amsmath,twocolumn,showpacs,amssymb,superscriptaddress]{revtex4} 
\pdfoutput=1
\usepackage{graphicx}
\usepackage{dcolumn}
\usepackage{bm}
\usepackage{stmaryrd}
\usepackage{latexsym}
\usepackage{amsfonts}

\begin{document}

\title{Nonequilibrium pairing instability in ultracold Fermi gases with population imbalance}

\author{Andrea Tomadin}
\email{a.tomadin@sns.it}
\affiliation{NEST-CNR-INFM and Scuola Normale Superiore, I-56126 Pisa, Italy}
\author{Marco Polini}
\affiliation{NEST-CNR-INFM and Scuola Normale Superiore, I-56126 Pisa, Italy}
\author{M.P. Tosi}
\affiliation{NEST-CNR-INFM and Scuola Normale Superiore, I-56126 Pisa, Italy}
\author{Rosario Fazio}
\affiliation{International School for Advanced Studies (SISSA), via Beirut 2-4, I-34014 Trieste, Italy}
\affiliation{NEST-CNR-INFM and Scuola Normale Superiore, I-56126 Pisa, Italy}

\date{\today}

\begin{abstract}
We present detailed numerical and analytical investigations of the nonequilibrium 
dynamics of spin-polarized ultracold Fermi gases following a sudden switching-on of the atom-atom 
pairing coupling strength. Within a time-dependent mean-field approach we show that on increasing 
the imbalance it takes longer for pairing to develop, the period of the nonlinear oscillations 
lengthens, and the maximum value of the pairing amplitude decreases. As expected, dynamical 
pairing is suppressed by the increase of the imbalance. Eventually, for a critical value of the 
imbalance the nonlinear oscillations do not even develop. Finally, we point out an interesting 
temperature-reentrant behavior of the exponent characterizing the initial instability. 
\end{abstract}

\pacs{03.75.Ss, 03.75.Kk.}

\maketitle

\section{Introduction}
\label{sect:intro}
One of the new exciting avenues that can be explored in the study of many-body properties of 
cold atomic gases~\cite{lewenstein_review,bloch_review} is the nonequilibrium dynamics following 
a sudden quench. Present-day technology allows to change the coupling constants~\cite{greiner_collapse} 
on such short time scales that it is possible to explore the regime where the many-body system is still 
governed by a unitary evolution but with nonequilibrium initial conditions. Time-dependent couplings 
can be realized, for example, by varying the intensity of the laser that fixes the amplitude of an optical 
lattice or by changing the atomic scattering length through sweeping an external magnetic field across a Feshbach 
resonance. This problem, which has attracted a lot of attention recently~\cite{altman02,BLS2004,AGR2004,
YAKE2005R,szymanska05,warner05,cazalilla06,YTA2006,BL2006,rigol_prl_2007,collath07,manmana07,cramer07,dzero07,
pasquale_prl_2006}, is what we consider as well.

Our work is inspired by Refs.~\onlinecite{BLS2004} and \onlinecite{YAKE2005R} that deal with the study of the dynamical 
pairing instability in cold atomic gases after a sudden switch of the attractive interaction  
at times shorter than the quasiparticle energy relaxation time. Barankov {\em et al}.~\cite{BLS2004}, starting 
from a normal state, showed that after the quench the system is unstable.  Pairing correlations initially 
build up exponentially in time and then oscillate taking the form of soliton trains. If the system before the 
quench is in an equilibrium BCS state, and the quench is performed by changing abruptly the pairing coupling, then
the stationary state can show a constant (but reduced) gap or can be gapless~\cite{YTA2006,YD2006}.
A classification of the allowed nonequilibrium behaviors arising from different initial conditions has been 
presented in Ref.~\onlinecite{YAKE2005}. To date there are no experiments on the non-adiabatic switching of 
pairing in fermion condensates. A proposal to detect signatures of nonequilibrium dynamics using 
radio-frequency spectroscopy has been put forward recently~\cite{dzero07}. 

Along the lines of these previous works (see also Ref.~\cite{galperin_1981}), in the present paper we study the pairing instability in 
a two-component ultracold Fermi gas with unbalanced spin populations after a sudden switch of the 
attractive interaction between the two fermion species. As is well known since the early days of 
superconductivity~\cite{Clogston1962,Sarma1963,FFLO}, an imbalance in the number densities of the two 
species tends to suppress pairing. Unbalanced Fermi gases~\cite{casalbuoni_2004} are currently attracting 
a great deal of experimental and theoretical interest. One of the aims is to detect exotic paired 
states~\cite{FFLO,muther02,liu03,bedaque03} that have been elusive so far in conventional solid-state systems. 
Fermi gases with population imbalance have been realized in a series of experiments~\cite{zwirlein06a,
partridge06,zwirlein06b,shin06}. The equilibrium phase diagram has been worked out in great detail (see, for example, 
Refs.~\onlinecite{mizushima05,sheehy06,pieri06,chien06,kinnunen06,PMLS2007} and references therein) and a very 
rich scenario has emerged.  However, despite the tremendous effort that has been devoted to understand 
equilibrium phases, nothing is known yet about the out-of-equilibrium properties of these system. Here 
we address this question for the first time. As a first step we analyze the instability of a normal partially 
spin-polarized Fermi gas with respect to s-wave pairing which leads to nontrivial results.  
Guided by the body of knowledge acquired in the study of the equilibrium case, one can look also for 
instabilities towards more complex paired states that we leave for future study. 

The time scales that are relevant to the present problem~\cite{BL2006} are the quasiparticle Landau
Fermi-liquid lifetime $\tau_{\rm el}$, the time $\tau_{\Delta}$ over which the oscillations of the pairing 
function develop and evolve~\cite{BLS2004}, and the characteristic time $\tau_{0}$ over which the coupling is 
switched on. We are interested in the regime when the inequalities $\tau_{0}\ll\tau_{\Delta}<t\ll\tau_{\rm el}$ hold.

The paper is organized as follows. In the next Section we first introduce the model 
Hamiltonian that we use to describe the system of interest. In 
Section~\ref{tdmean} we discuss the mean-field decoupling used to study the time evolution, while 
in Sect.~\ref{sect:initial_state} we carefully describe the initial state to which the quench is applied. 
The resulting equations can be analyzed 
both numerically and analytically. In Sect.~\ref{sect:numerical_simulations} 
we present our numerical simulations of the time-dependent mean-field equations and discuss their 
main features. In Sects.~\ref{sect:linear_instab} and~\ref{sect:pairing_oscill} we present some analytical 
results for the short- and 
long-time properties of the quantum evolution. In Sect.~\ref{sect:conclusions} we summarize 
our main conclusions. Finally, Appendix~\ref{appendix:app_1} contains more details on the 
numerical simulations of the time-dependent mean-field equations, 
while Appendices~\ref{appendix:app_2} and~\ref{appendix:app_3} contain some details 
of the calculations presented in Sect.~\ref{sect:linear_instab}.

\section{The model}
\label{sect:model}

The time-dependent BCS Hamiltonian is defined as
\begin{equation}
\label{eq:model1}
        \hat{\cal H}_{\rm BCS}(t)
	= \sum_{{\bm k},\sigma}\varepsilon_{{\bm k}}\hat{c}_{{\bm k}\sigma}^{\dagger}\hat{c}_{{\bm k}\sigma}
	+ g(t)\sum_{{\bm k},{\bm k}'}
	\hat{c}_{{\bm k}\uparrow}^{\dagger}\hat{c}_{-{\bm k}\downarrow}^{\dagger}
	\hat{c}_{-{\bm k}'\downarrow}\hat{c}_{{\bm k}'\uparrow}~.
\end{equation}
In this equation $\hat{c}_{{\bm k}\sigma}^{\dagger}$ ($\hat{c}_{{\bm k}\sigma}$)  creates (annihilates) a 
fermion with momentum ${\bm k}$ ($\hbar=1$) and spin $\sigma=\uparrow,\downarrow$
(hyperfine state label). The number $N_\sigma$ of particles with spin $\sigma$ 
is fixed during the time evolution and thus we do not need to introduce chemical potentials 
for each spin species~\cite{Bulgac1990}. 
Given $N_\sigma$, the equilibrium Fermi energies $\varepsilon_{{\rm F}\uparrow}$ and 
$\varepsilon_{{\rm F}\downarrow}$ of the noninteracting system at zero temperature are fixed.
The summations in Eq.~(\ref{eq:model1}) are carried out over a shell of energies of thickness 
$2\omega_{\rm D}$ around the Fermi energies, where $\omega_{\rm D}$ is an effective ultraviolet 
cutoff frequency \cite{footnote1}. We assume that the Fermi energy 
mismatch, $\delta\mu \equiv \varepsilon_{{\rm F}\uparrow}-\varepsilon_{{\rm F}\downarrow}$, is smaller than $\omega_{\rm D}$. 
For convenience we measure all the energies from $\varepsilon_{{\rm F}\downarrow}$ 
and approximate the parabolic dispersion $\varepsilon_{\bm k}$ with a sequence of $N\gg 1$ equally spaced levels $\varepsilon_{k}$ in 
the range $[-\omega_{\rm D},\omega_{\rm D}]$, where $k=1\dots N$ is a scalar label.
The level spacing is $\delta\varepsilon=2\omega_{\rm D}/(N-1)$ 
and the density of states is $1/\delta\varepsilon$. 

The coupling $g(t)$ is zero if $t\le 0$  and is switched on to a constant negative value $-g$ 
during a time interval  $0<t \lesssim \tau_{0}$. Since we focus on the non-adiabatic evolution  
($t_{0}\ll\tau_{\Delta}$), we approximate $g(t)\approx -g\Theta(t)$, where $\Theta(x)$ is the Heaviside 
step function. It is worth to notice that if the switching on of the interaction is too fast, the gas 
becomes overheated and the time-dependent coupling induces two-particle scattering. However, as 
discussed in Ref.~\onlinecite{BL2006}, a time window for $\tau_{0}$ exists in which the constraint 
for avoiding the overheating is compatible with that of a sudden switching-on of the interaction.

\subsection{Time-dependent mean-field theory}
\label{tdmean}

As discussed in Ref.~\onlinecite{BL2006}, the nonequilibrium evolution of the fermion system 
can be analyzed within a time-dependent mean-field theory. To this end we introduce the pairing function 
$ \Delta(t)=g\sum_{k}\langle \hat{c}_{-k\downarrow} \hat{c}_{k\uparrow} \rangle $,
where the average is taken over the quantum state of the system at time $t$.
After the mean-field decoupling is performed, the BCS Hamiltonian (\ref{eq:model1}) 
reduces to a sum of time-dependent commuting terms $\hat{\cal H}_{\rm MF}(t)=\sum_{k}\hat{\cal H}_{\rm MF}^{(k)}(t)$, where 
\begin{equation}\label{eq:meanfield}
\hat{\cal H}_{\rm MF}^{(k)}(t)
= \sum_{\sigma}\varepsilon_{k}\hat{c}_{k\sigma}^{\dagger}\hat{c}_{k\sigma}
	- \Delta(t) \hat{c}_{k\uparrow}^{\dagger}\hat{c}_{-k\downarrow}^{\dagger}
	- \Delta^{\ast}(t) \hat{c}_{-k\downarrow}\hat{c}_{k\uparrow}~.
\end{equation}
Within the mean-field approximation the Hilbert space to study the time evolution of the system is 
the tensor product of $N$ Fock spaces with at most two particles instead of the larger Fock space 
with at most $2N$ particles. 
There are only four states in the two-particle Fock space built with the single-particle orbitals: the 
vacuum state $|0\rangle$, a fully-occupied state $|2\rangle$ with two particles, and two singly-occupied 
states $|\!\!\uparrow\rangle$ and $|\!\!\downarrow\rangle$ labeled by the spin of each unpaired fermion.
Writing the Fock basis in this order, the matrix $\hat{\cal H}_{\rm MF}(t)$ within a block with a given $k$ is
\begin{equation}\label{eq:hspace6}
\hat{\cal H}^{(k)}_{\rm MF}(t)=
\left (
\begin{array}{cccc}
0	&-\Delta^{*}(t)	& 0	& 0\\
-\Delta(t)	&	2\varepsilon_{k} & 0	& 0 \\
 0	& 0	& \varepsilon_{k}	& 0		\\
 0	& 0	& 0	& \varepsilon_{k}	\\
\end{array}
\right )~.
\end{equation}
The Hamiltonian decomposes into four blocks along the diagonal.
The last two blocks are one-dimensional and determine the free evolution of the unpaired states, 
as these states cannot be coupled to the $|0\rangle \oplus |2\rangle $ condensate sector due to the 
Pauli-blocking effect. The two-dimensional block represents a Cooper pair, where the vacuum 
$|0\rangle$ is coherently coupled to the doubly-occupied state $|2\rangle$. The coupling is due to 
the pairing term $\hat{c}_{k\uparrow}^{\dagger}\hat{c}_{-k\downarrow}^{\dagger}$ that does not conserve 
the number of particles within the subspace.

Since it is important to include the case where the fermions can be excited out of the condensate 
into unpaired states by incoherent thermal processes, a wave function is not appropriate to treat 
the evolution of the two-particle system. To treat this problem we use a statistical matrix defined as
\begin{eqnarray}
\label{eq:stat_mat}
    \rho^{(k)}(t) & = &
	(1-p_{k\uparrow}-p_{-k\downarrow})
	[\tilde{u}_{k}(t)|0\rangle + \tilde{v}_{k}(t)|2\rangle ] \nonumber \\ 
	 & \mbox{} \times  & [\langle 0|\tilde{u}_{k}^{\ast}(t) + \langle 2|\tilde{v}_{k}^{\ast}(t)]
    +  p_{k\uparrow} |\!\!\uparrow \rangle \langle\uparrow\!\!| 
	+ p_{-k\downarrow}|\!\!\downarrow\rangle \langle\downarrow\!\!|~. \nonumber \\
\end{eqnarray}
The probabilities $p_{k\uparrow}$ and $p_{-k\downarrow}$ take into account the thermal excitation of 
particles out of the condensate. We remark that each pure state that enters the construction 
of the statistical matrix has to be normalized, {\it i.e.} $|\tilde{u}_{k}(t)|^{2}+|\tilde{v}_{k}(t)|^{2}=1$.

Both the Hamiltonian (\ref{eq:hspace6}) and the statistical matrix (\ref{eq:stat_mat}) are block-diagonal 
and the condensate sector evolves independently of the other states, according to 
$i\partial_{t}\rho^{(k)}(t)=[\hat{\cal H}_{\rm MF}^{(k)}(t), \rho^{(k)}(t)]$.
We can define an effective Hamiltonian $\hat{\cal H}_{\rm c}^{(k)}$ restricted to the condensate sector 
and an effective state vector $|\Xi_{k}(t)\rangle=u_{k}(t)|0\rangle + v_{k}(t)|2\rangle $, with 
$u_{k}(t)=\tilde{u}_{k}(t)(1-p_{k\uparrow}-p_{-k\downarrow})^{1/2}$ and $v_{k}(t)=\tilde{v}_{k}(t)(1-p_{k\uparrow}-p_{-k\downarrow})^{1/2}$.
The statistical matrix projected onto the condensate sector then reads 
$\rho^{(k)}_{\rm c}(t)=|\Xi_{k}(t)\rangle \langle \Xi_{k}(t)|$.  The pure-state form of the projected statistical matrix 
is preserved by the time evolution.
This implies that the effective, non-normalized state vector $|\Xi_{k}(t)\rangle$ belonging to the condensate sector 
$|0\rangle \oplus |2\rangle $ is sufficient to describe the time evolution.

The state vector $|\Xi_{k}(t)\rangle$ evolves according to the norm-preserving effective Schr\"odinger equation 
$ i\partial_{t}|\Xi_{k}(t)\rangle =\hat{\cal H}_{\rm c}^{(k)}(t)|\Xi_{k}(t)\rangle$, and the coefficients 
$u_{k}(t)$ and $v_{k}(t)$ obey the time-dependent Bogolubov--de Gennes equations (BdGE)
\begin{equation}\label{eq:bdge}
i\partial_{t}
	\left(\begin{array}{c}
	v_{k}(t) \\
	u_{k}(t)
	\end{array} \right) =
	\left( \begin{array}{cc}
	\varepsilon_{k} & -\Delta(t) \\
	-\Delta^{\ast}(t) & -\varepsilon_{k}
	\end{array} \right)	
	\left(\begin{array}{c}
	v_{k}(t) \\
	u_{k}(t)
	\end{array} \right)~.
\end{equation}
The total Fock space for (at most) $2N$ particles is then defined to be the tensor product of the 
two-particle spaces and the statistical matrix is $\rho=\bigotimes_{k}\rho^{(k)}$.
If an operator $\hat{\cal O}_{k}$ has support within the condensate sector of the $k$ space, its 
expectation value ${\rm Tr}[\rho\hat{\cal O}_{k}]$ can be computed using the effective state vector 
only and reads $\langle\Xi_{k}(t)|\hat{\cal O}_{k}|\Xi_{k}(t)\rangle$.
The BdGE have to be solved together with the self-consistency condition
\begin{equation}
\Delta(t) = g\sum_{k}u_{k}^{*}(t)v_{k}(t)~.
\end{equation}

\subsection{The initial state}
\label{sect:initial_state}

The BdGE in~(\ref{eq:bdge}) must be accompanied by some initial conditions $U_{k}=u_{k}(t=0)$ 
and $V_{k}=v_{k}(t=0)$.
The initial conditions thus describe the state of the system just before the quench is applied at time 
$t \to 0^+$. 
We have chosen initial conditions corresponding to the equilibrium  
configuration of the Hamiltonian~(\ref{eq:meanfield}) 
at temperature $\theta$ and $g=0$. 
We compute the partition function ${\cal Z}_{k}$ of the $k$-th subsystem 
in the grand-canonical ensemble
\begin{equation}
{\cal Z}_{k}=1+e^{-\beta(2 \varepsilon_{k} - \mu_{\uparrow} - \mu_{ 
\downarrow})} + e^{-\beta(\varepsilon_{k}-\mu_{\uparrow})} + e^{- 
\beta(\varepsilon_{k}-\mu_{\downarrow})}~,
\end{equation}
where $\beta=1/\theta$ ($k_{\rm B}=1$), $\mu_{\uparrow}$ and $\mu_{\downarrow}$ are the chemical potentials
for the two spin species, and the difference $\mu_{\uparrow}-\mu_{\downarrow}$ 
is equal to the Fermi energy mismatch $\delta\mu$.

The probability to find the system in the state $|2 \rangle$ is
\begin{eqnarray} 
|V_{k}|^{2}&=&\langle 2|\rho^{(k)}|2\rangle=\frac{1}{{\cal Z}_k}\exp[-\beta(2 \varepsilon_{k} - \mu_{\uparrow} 
- \mu_{\downarrow})]\nonumber\\
&=&f_{k\uparrow}f_{k\downarrow}~,
\end{eqnarray} 
with $f_{k\uparrow}=\{1+{\rm exp}[\beta(\varepsilon_{k}-\delta\mu)]\}^{-1}$ 
and $f_{k\downarrow}=[1+{\rm exp}(\beta\varepsilon_{k})]^{-1}$. 
Similarly, the probability to find the system in the state $|0\rangle$ is 
$|U_{k}|^{2}=(1-f_{k\uparrow})(1-f_{k\downarrow})$.
The probability $|U_{k}|^{2} + |V_{k}|^{2}$ to find the $k$-th subsystem  
in the condensate sector is smaller than unity: because of thermal excitations 
there is a finite probability that the  $k$-th 
subsystem is occupied by an unpaired fermion. It is easy to see that the
expression $|u_{k}(t)|^{2}+|v_{k}(t)|^{2}$ is constant in time.

Since at times $t\leq 0$ the system is noninteracting, 
the phase $\phi_{k}$ of the coherence $\langle 2|\rho^{(k)}|0\rangle=U^{\ast}_{k} V_{k}$ is a random variable  
of $k$. As a consequence we can take as initial conditions
\begin{eqnarray}\label{eq:initial_conditions}
U_{k} & = & \sqrt{1-f_{k\uparrow}}\sqrt{1-f_{k\downarrow}} 
\nonumber \\
V_{k} & = & \exp{[i\phi_{k}]}~\sqrt{f_{k\uparrow}f_{k\downarrow}}~.
\end{eqnarray}
A non-zero temperature or a finite value of the imbalance are sufficient to produce a non-zero initial
pairing amplitude $|\Delta(t=0)|$, which is very small because of the randomness of the initial phases. 

\section{Results}
\label{sect:results}

In this Section we discuss our results for the time dependence of the pairing $\Delta(t)$ and the 
distribution of paired particles $n_k(t)$ as functions of spin imbalance, temperature, and initial 
conditions. The numerical results, obtained through integration of the BdGE, will be supplemented 
by analytical results obtained in the short-time and stationary regimes.

\subsection{Numerical solution of the BdGE}
\label{sect:numerical_simulations}

We now turn to the presentation of the numerical solution of the BdGE (\ref{eq:bdge}) 
with initial conditions given in Eq.~(\ref{eq:initial_conditions}).
In what follows we use as unit of energy the real quantity $\Delta_{0}$ 
defined by the solution of the equilibrium 
BCS self-consistency equation $g\sum_{k}(\varepsilon_{k}^{2}+\Delta_{0}^{2})^{-1/2}$=\,2.
This choice of the energy scale then fixes the value of $g$.
Frequency and time scales are defined accordingly. To solve the BdGE we have used a fourth-order 
adaptive-stepsize Runge-Kutta algorithm, with a maximum relative error of $10^{-5}$ per time step. 
A typical time step is $10^{-3}-10^{-2}$, but a smaller time step 
of order $10^{-5}$ is used near the initial instability of the BdGE (see below).
The integration of the BdGE up to $t_{\rm max}=300$ takes less than $10~{\rm secs}$ on a desk PC. 

\begin{figure}[t]
\begin{center}
\includegraphics[width=1.00\linewidth]{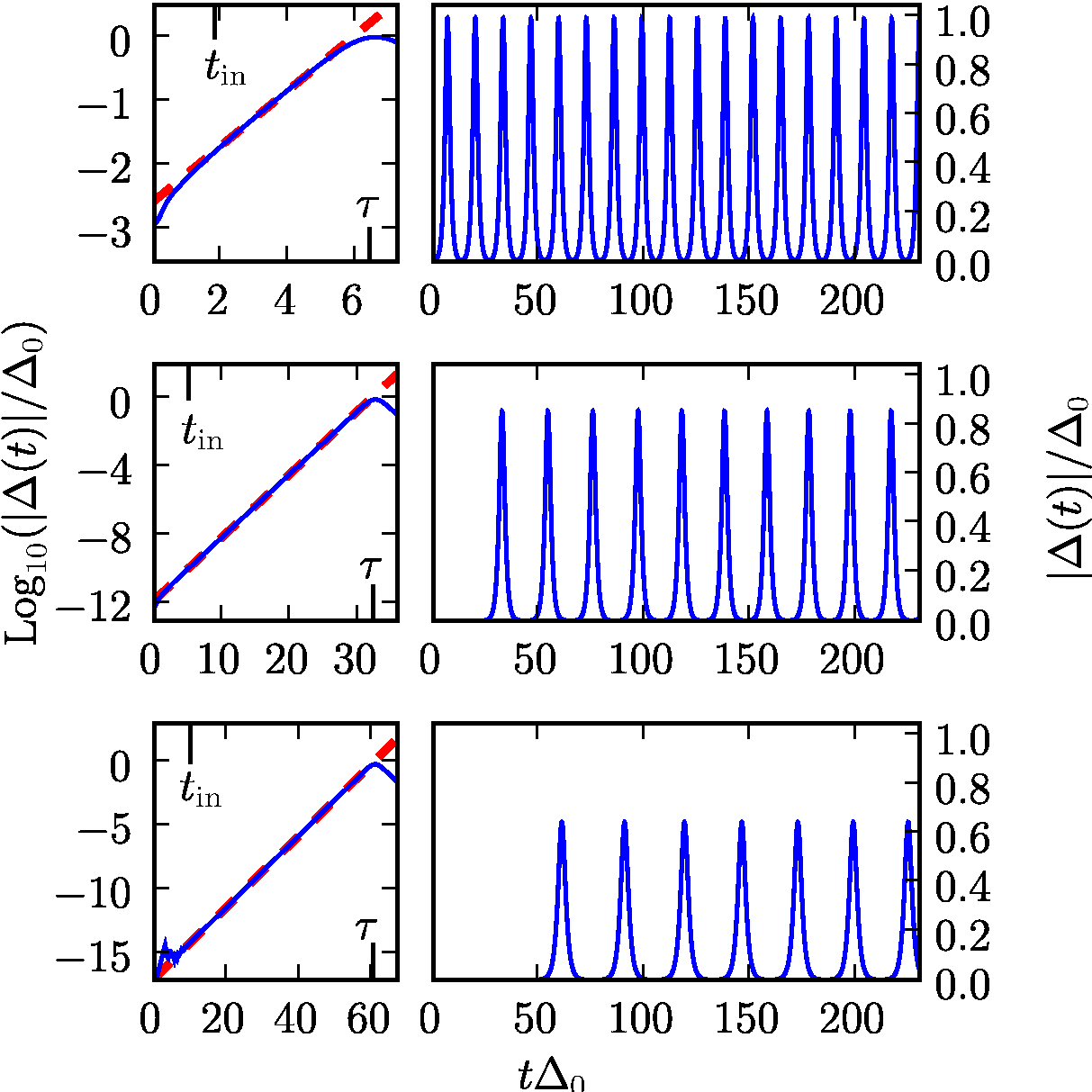}
\caption{(Color online) Modulus $|\Delta(t)|$ of the pairing function (in units of $\Delta_{0}$) as a function of time (in units of $1/\Delta_{0}$), 
obtained by solving the BdGE. 
From top to bottom the value of the initial Fermi-energies mismatch increases as 
$\delta\mu=0.0, 0.5,$ and $0.75$. The left panels show 
a zoom of the initial linear instability region in the range $t<\tau$, $\tau$ being the time 
at which $|\Delta(t)|$ has its first peak. The time interval $[0,t_{\rm in}]$ is the transient discussed in (i) in Sect.~\ref{sect:numerical_simulations}.
The thick dashed lines are linear fits in the range $0.3\tau<t<0.8\tau$: the slope of each dashed 
line gives $\gamma$, while the extrapolation to $t=0$ gives $\eta$. 
A computer precision of $10^{-15}$ is reached for $\delta\mu=0.75$ (bottom panel) and $t<10.0$, where 
fluctuations due to the numerics begin to appear.\label{fig:one}}
\end{center}
\end{figure}

In Fig.~\ref{fig:one} we show some representative results of the solution of the BdGE for
$N=10^{3}$, $\omega_{\rm D}=5.0$ and $g\simeq 4 \times 10^{-3}$.
We choose three initial states with different imbalance $\delta\mu$ at a temperature $\theta=10^{-2}$.
Each profile is obtained with a random realization of the initial phases $\phi_{k}$ that we take as 
uniformly distributed in the interval $[0,2\pi]$.

Three time regimes are evident for each value of the imbalance $\delta \mu$ in 
Fig.~\ref{fig:one}: (i) a very short initial transient $[0,t_{\rm in}]$ where the pairing amplitude 
increases by several orders of magnitude, as will be clarified in Sect.~\ref{sect:linear_instab}; (ii) a time interval $[t_{\rm in},\tau]$ in which the growth of $|\Delta(t)|$ is exponential in time, $|\Delta(t)|=\eta\exp{(\gamma t)}$ 
($\tau$ will be hereafter referred to as ``time lag'', following the jargon introduced in Ref.~\onlinecite{BLS2004}); 
and (iii) a time interval where undamped, nonlinear oscillations of $|\Delta(t)|$ occur. 

Several observations are in order at this point. 
On increasing the imbalance $\delta\mu$ the exponent $\gamma$ of the exponential growth 
in region (ii) decreases and the time lag $\tau$ increases ({\it i.e.} it takes longer for 
pairing to develop), the period of the nonlinear oscillations lengthens, and the maximum value 
of the pairing amplitude decreases. As expected, dynamical pairing is suppressed by the increase 
of the imbalance. 
In Sect.~\ref{sect:linear_instab} we prove that dynamical pairing is wholly suppressed at a 
critical value $\delta\mu_{\rm c}$ of the imbalance.
It is hard to verify this assertion numerically because at large imbalance the initial pairing 
$|\Delta(t=0)|$ becomes comparable to the computer accuracy. 

To test the robustness of the profiles shown in Fig.~\ref{fig:one} against changes in the initial 
conditions we have solved the BdGE with several different choices of the initial random phases.
The results of this statistical analysis are reported 
in Appendix~\ref{appendix:app_1}, where we show that the amplitude of the pairing 
is essentially independent of the particular realization of the random phases. 

\begin{figure}[t]
\begin{center}
\includegraphics[width=1.0\linewidth]{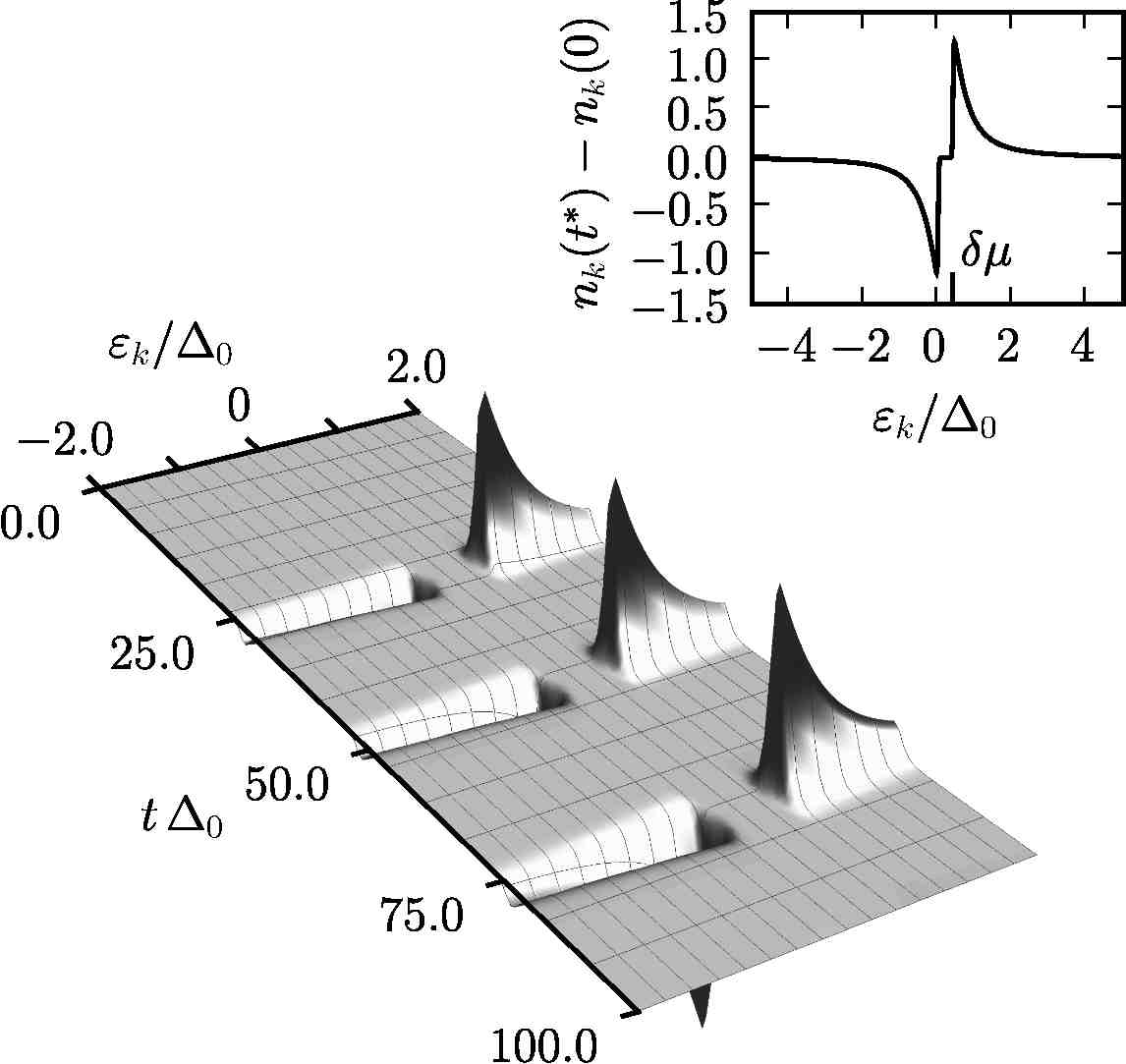}
\caption{A three-dimensional plot of the quantity $n_k(t)-n_k(0)$ 
as a function of energy $\varepsilon_{k}$ and of time $t$. In the top-right panel we show $n_k(t^\ast)-n_k(0)$ as a function of $\varepsilon_k$ 
at a time instant $t^{\ast}$ where the pairing amplitude is maximal. In this figure $\delta\mu=0.5$, as in
the central panel in Fig.~\ref{fig:one}.\label{fig:two}}
\end{center}
\end{figure}

In Fig.~\ref{fig:two} we show the distribution of condensed particles
\begin{equation}\label{eq:nkappa}
n_k(t)=\sum_\sigma \langle \Xi_{k}(t) |  {\hat c}^\dagger_{k\sigma}{\hat c}_{k\sigma} | \Xi_{k}(t) \rangle=2|v_k(t)|^2~,
\end{equation}
measured from its initial value $n_{k}(0)$, as a function of energy $\varepsilon_{k}$ and time $t$. 
As a function  of time, the quantity $n_k(t)$ is always nearly equal to its initial value $n_k(0)$ except 
in close proximity to the maxima of the pairing amplitude $|\Delta(t)|$. 
As time evolves, $n_{k}(t)-n_{k}(0)$ pulses in synchronism with the nonlinear oscillations of the pairing function.
Close to a time $t^\ast$ at which the pairing amplitude is maximal, $n_k(t)-n_k(0)$
exhibits a peculiar structure (see top-right panel in Fig.~\ref{fig:two}). 
We in fact see a downward peak in the region below the Fermi surface of the 
minority-spin component and an upward peak, equal in size to the downward one, located 
above the Fermi surface of the majority-spin component.
In between the two peaks we recognize a region of extension $\delta\mu$ where pairing is suppressed
since the condensate sectors $|0\rangle\oplus|2\rangle$ are almost entirely depleted, 
{\it i.e.} $|u_{k}(t^\ast)|^{2}+|v_{k}(t^\ast)|^{2}\simeq 0$ for $\varepsilon_k \in [0,\delta\mu]$.
The two peaks indicate that particles in the condensate are transferred across the Fermi surfaces of the two populations.
This phenomenon is reminiscent of what happens in conventional BCS equilibrium superconductivity. 

In what follows, we show that the different regimes of the initial onset of the pairing 
instability and of the nonlinear oscillations 
are amenable to an analytical treatment. In particular, in Sect.~\ref{sect:linear_instab} 
we solve by means of a linear-stability analysis the time-dependent BdGE in 
the time interval $[0,\tau]$ (regions (i) and (ii) introduced above). 
In Sect.~\ref{sect:pairing_oscill} we discuss the stationary limit within the general theoretical 
framework that was earlier developed in Refs.~\onlinecite{YTA2006} and~\onlinecite{YAKE2005} for the unpolarized case.
The main result of these two sections is a complete analytical prediction of the solutions of the 
time-dependent BdGE.

\subsection{Analysis of linear instability}
\label{sect:linear_instab}

The initial build-up of the pairing instability can be studied by means of a linear-stability analysis,
along the lines of what was earlier done in Ref.~\onlinecite{BLS2004} for the unpolarized case. 

It is convenient to  introduce the following definitions, corresponding to a free evolution of each 
Cooper pair,
\begin{eqnarray}\label{eq:linz1}
\bar{u}_{k}(t) & = &	e^{+i\varepsilon_{k}t}U_{k} \nonumber \\
\bar{v}_{k}(t) & = &	e^{-i\varepsilon_{k}t}V_{k} \nonumber \\
\bar{\Delta}(t) & = &	g\sum_{k}\bar{u}_{k}^{\ast}(t)\bar{v}_{k}(t) 
= g\sum_{k}U_{k}^{\ast}V_{k}e^{-i2\varepsilon_{k}t}~,
\end{eqnarray}
where $U_k$ and $V_k$ are the initial values in Eq.~(\ref{eq:initial_conditions}). 
Without loss of generality, we can write any solution of the BdGE in the form 
$u_{k}(t) =  \bar{u}_{k}(t)+\delta u_{k}(t)$ and $v_{k}(t) = \bar{v}_{k}(t)+\delta v_{k}(t)$.
We choose $\delta u_{k}(0)=\delta v_{k}(0)=0$ so that 
the initial conditions are still given by $u_{k}(0)=U_{k}$ and $v_{k}(0)=V_{k}$.
Inserting these definitions into the BdGE 
we obtain the equations of motion for the corrections $\delta u_{k}(t)$ and $\delta v_{k}(t)$,
\begin{equation}\label{eq:linz8}
i\partial_{t} 
\left(\begin{array}{c}
	\delta u_{k}(t) \\
	\delta v_{k}(t)
\end{array} \right) = 
\left(\begin{array}{c}
	-\varepsilon_{k}\delta u_{k}(t) - \Delta^{\ast}(t)[\bar{v}_{k}(t)+\delta v_{k}(t)] \\
	-\Delta (t) [\bar{u}_{k}(t)+\delta u_{k}(t)]+\varepsilon_{k} \delta v_{k}(t)
\end{array}\right)~.
\end{equation}
We solve Eq.~(\ref{eq:linz8}) in a time interval $t_{\rm in} < t \lesssim \tau$ 
defined by the hypotheses
\begin{eqnarray}
& \mbox{(i) } & |\Delta(t)|\gg|\bar{\Delta}(t)| \nonumber \\
& \mbox{(ii) } & |\delta v_{k}(t)|\ll|\bar{v}_{k}(t)|~.
\end{eqnarray}
These hypotheses mean that after an ``instability time'' $t_{\rm in}$ the pairing function 
built up by the corrections $\delta u_{k}$ and $\delta v_{k}$ is much larger than the pairing due 
to the unperturbed functions $\bar{u}_{k}$ and $\bar{v}_{k}$.
The first hypothesis is fulfilled if the initial state is \emph{weakly paired}, {\it i.e.} if 
$\bar{\Delta}(0) = g|\sum_{k}U_{k}^{\ast}V_{k}| \ll \Delta_{0}$. 
The second hypothesis guarantees that the corrections are much smaller than the unperturbed 
functions, so that we can neglect the nonlinear terms $\delta u_{k}^{\ast}\delta v_{k}$ in the pairing function. 
The nonlinear terms become important only after a ``nonlinearity time'' $\lesssim \tau$.

The time evolution in the interval $[t_{\rm in},\tau]$ is ruled by the linear ordinary 
differential equation
\begin{equation}\label{eq:linz12}
i\partial_{t}
	\left(\begin{array}{c}
		\delta u_{k}(t)\\
		\delta v_{k}(t)
	\end{array}\right) = 
	\left(\begin{array}{c}
		-\varepsilon_{k}\delta u_{k}(t)-\delta\Delta^{\ast}(t)\bar{v}_{k}(t) \\
		-\delta \Delta(t)\bar{u}_{k}(t)+\varepsilon_{k}\delta v_{k}(t)
	\end{array}\right)~,
	\end{equation}
where $\delta \Delta(t) \equiv g\sum_{k}[\bar{u}_{k}^{\ast}(t) \delta v_{k}(t) + \delta u_{k}^{\ast}(t)\bar{v}_{k}(t)]$. 
This equation does not allow us to trace the nonlinear evolution in the interval $[0,t_{\rm in}]$. 
We only need to assume that $\delta u_{k}(t_{\rm in})$, $\delta v_{k}(t_{\rm in})$ and $\delta\Delta(t_{\rm in})$ are non-zero 
and we write the following {\it Ansatz} for the solution of Eq.~(\ref{eq:linz12}) at times $t> t_{\rm in}$:
\begin{eqnarray}\label{eq:linz14}
\delta \Delta(t) & = & e^{-i\zeta (t-t_{\rm in})}~\delta \Delta(t_{\rm in}) \nonumber \\
\delta  u_{k}(t) & = & e^{-i(\varepsilon_{k}-\zeta^{\ast})(t-t_{\rm in})}\delta u_{k}(t_{\rm in}) \nonumber \\
\delta  v_{k}(t) & = & e^{+i(\varepsilon_{k}-\zeta)(t-t_{\rm in})}\delta v_{k}(t_{\rm in})~.
\end{eqnarray}
Here we have introduced a complex instability exponent $\zeta=\omega+i\gamma$.
Inserting the {\it Ansatz} (\ref{eq:linz14}) in Eq.~(\ref{eq:linz12}) one can easily obtain~\cite{BL2006} the following 
``consistency relation'' for the instability exponent $\zeta$,
\begin{equation}\label{eq:linz20}
\sum_{k}\frac{|U_{k}|^{2}-|V_{k}|^{2}}{2\varepsilon_{k}-\zeta} - \frac{1}{g} = 0~.
\end{equation}
This equation is identical in form to Eq.~(18) in Ref.~\onlinecite{BL2006}, but here the solution $\zeta=\zeta(\delta\mu,\theta)$ depends on two physical parameters: the imbalance $\delta\mu$ and the temperature $\theta$ (rather than only on temperature, as in the unpolarized case). 
For $\delta\mu=0$ we recover the results in Fig.~10 of Ref.~\onlinecite{BL2006}. 

\begin{figure}
\begin{center}
\includegraphics[width=0.80\linewidth]{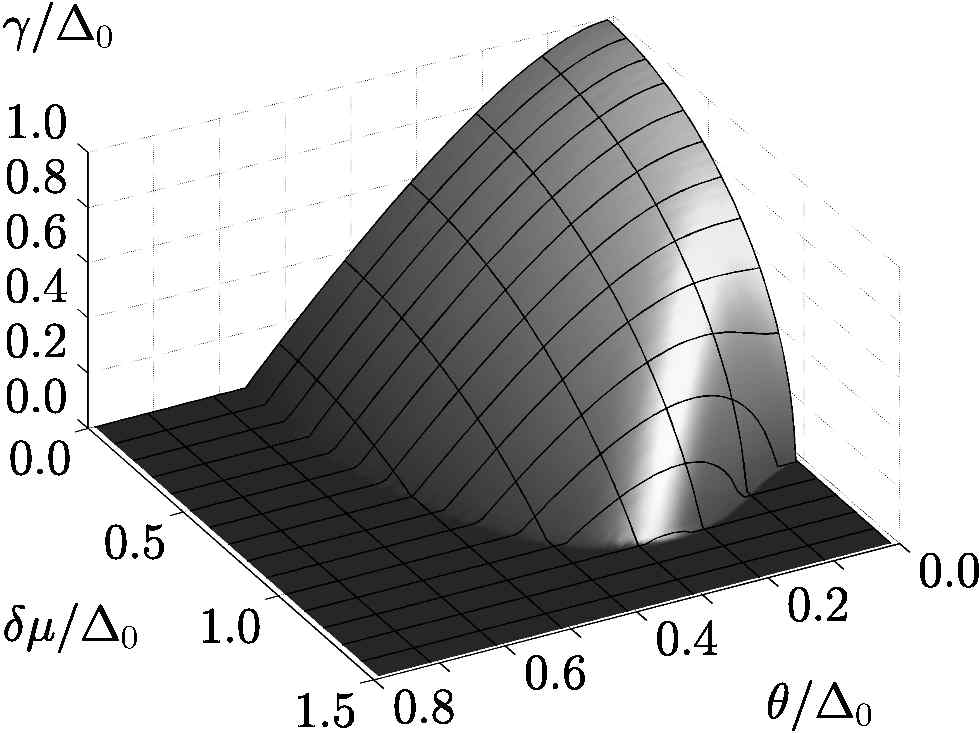}
\vspace{0.4in}\\
\includegraphics[width=0.85\linewidth]{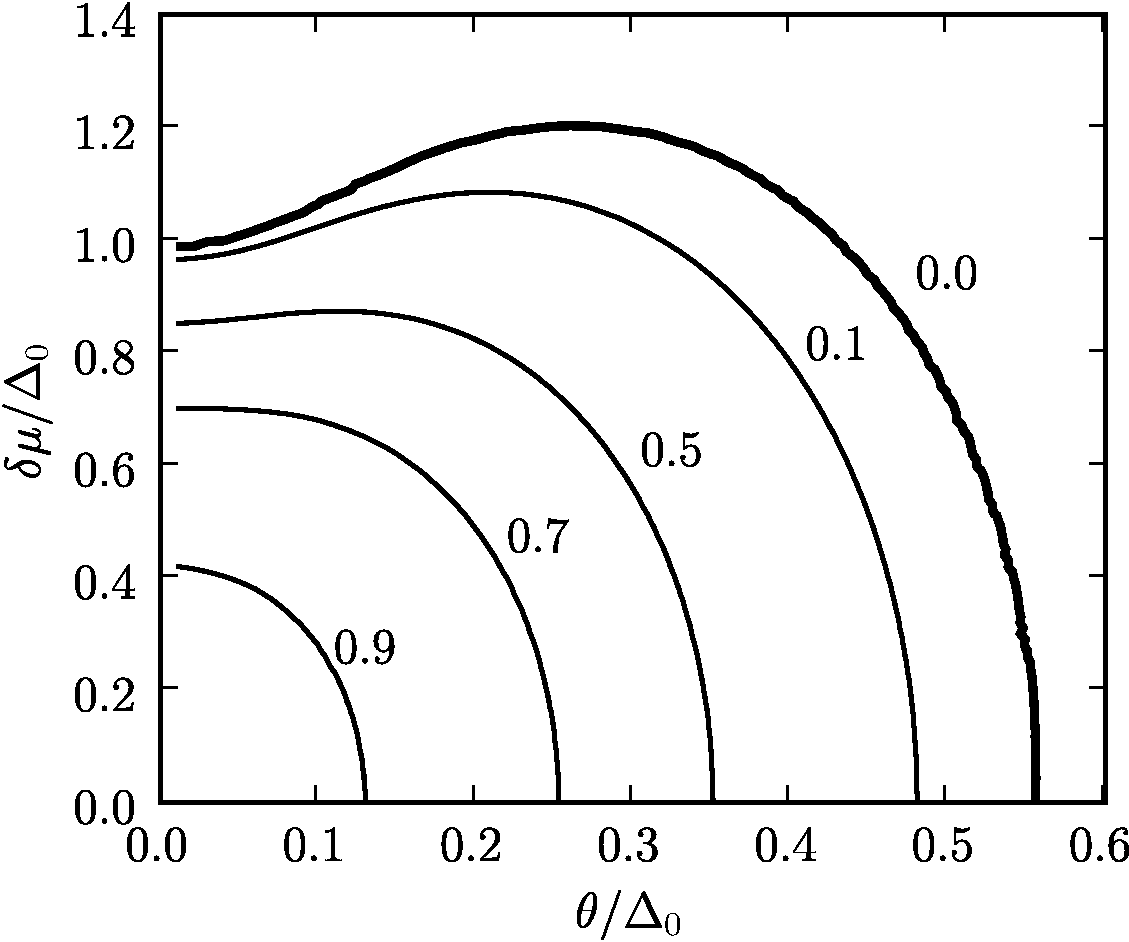}
\caption{Top panel: a plot of the imaginary part $\gamma=\gamma(\delta\mu,\theta)$ 
of the instability exponent $\zeta$ as a function of $\delta\mu$ and $\theta$. 
Bottom panel: contour plots corresponding to the top panel. The thick solid line shows the points 
of the $(\delta\mu,\theta)$ plane where the 3D profile in the top panel intersects 
the $\gamma=0$ plane (in actuality this curve has been calculated for $\gamma=10^{-2}$ for numerical 
reasons). The re-entrance described in the main body of the text is clearly visible.\label{fig:three}}
\end{center}
\end{figure}

In Fig.~\ref{fig:three} we show the imaginary part of the solution of Eq.~(\ref{eq:linz20}) 
in the $(\delta\mu,\theta)$ plane. To solve Eq.~(\ref{eq:linz20}) we have minimized the square of the l.h.s. 
with respect to the two parameters $\omega$ and $\gamma$. The minimum of the square is just the value 
where the l.h.s. vanishes. Several observations need to be done on Fig.~\ref{fig:three}. To begin with,
there is a critical line in the $(\delta\mu,\theta)$ plane  above which no instability develops, {\it i.e.} $\gamma=0$. 
The imaginary part $\gamma$ of the instability exponent decreases monotonically as a function of $\delta\mu$. 
On the contrary, $\gamma$ depends monotonically on temperature only if $\delta\mu < \delta\mu_{\rm r}\simeq 0.7$.
In this case $\gamma$ decreases if $\theta$ increases, while the opposite behavior happens if $\delta\mu > \delta\mu_{\rm r}$ 
and the temperature is low. The latter region of the $\delta\mu-\theta$ plane appears as a re-entrance in the bottom panel of Fig.~\ref{fig:three}. 
In this region an increase in temperature allows the system to sustain pairing even in the presence of 
a larger maximum imbalance. This is reminiscent of a similar re-entrant behavior obtained in the equilibrium case
by Sarma~\cite{Sarma1963}. In that case, however, the author found the existence of a more stable phase characterized by 
the absence of re-entrance. The calculations in Ref.~\onlinecite{Sarma1963} are equilibrium calculations 
performed within a grand-canonical ensemble and thus do not rule out the possibility of a 
re-entrance in the ``phase diagram'' of Fig.~\ref{fig:three} for the out-of-equilibrium dynamics.

In some limiting cases it is possible to extract analytically the solution of Eq.~(\ref{eq:linz20}). 
In the thermodynamic limit, defined by letting $N\to \infty$ while keeping $\Delta_{0}$ and $\omega_{\rm D}$ fixed, 
Eq.~(\ref{eq:linz20}) reduces to~\cite{footnote}
\begin{eqnarray}\label{eq:imba5}
	&&\int_{-\omega_{\rm D}}^{\omega_{\rm D}}d\varepsilon~\frac{2\varepsilon-\omega}
	{(2\varepsilon-\omega)^{2}+\gamma^{2}} 
	[1-f_{\uparrow}(\varepsilon)-f_{\downarrow}(\varepsilon)] 
	-\frac{\delta\varepsilon}{g}   =  0 \nonumber \\
	&&\int_{-\omega_{\rm D}}^{\omega_{\rm D}}d\varepsilon~\frac{1}
	{(2\varepsilon-\omega)^{2}+\gamma^{2}}
	[1-f_{\uparrow}(\varepsilon)-f_{\downarrow}(\varepsilon)]
	  =  0~,
\end{eqnarray}
where the real and the imaginary part have been written separately. 
The Fermi functions $f_{\sigma}(\varepsilon)$ weigh the states that take part in the pairing process. 
The states in which there is a high probability 
to find an unpaired electron are effectively removed from the system.
This is most clearly seen at $\theta=0$, where the Fermi functions become sharp steps and 
$1-f_{\uparrow}(\varepsilon)-f_{\downarrow}(\varepsilon) 
= \Theta(\varepsilon-\delta\mu) - \Theta(-\varepsilon)$, thus 
excluding the interval $[0,\delta\mu]$ from the integrations in Eq.~(\ref{eq:imba5}). We see that the exclusion of some fermions from the pairing must lead to a decrease in the exponent $\gamma$ of the instability, or equivalently in the maximum amplitude $\Delta_{+}$ of the oscillations.
In the zero temperature $\theta=0$ case (see Appendix~\ref{appendix:app_2}), after performing an asymptotic expansion in powers of $1/\omega_{\rm D}$ we find that the solution of Eqs.~(\ref{eq:imba5}) is
\begin{equation}\label{eq:tzeroth4}
\gamma(\theta=0,\delta\mu) = \sqrt{1-\delta\mu^{2}}~,
\end{equation}
for $0<\delta\mu<1$. We thus see how an imbalance larger than $\delta\mu_{\rm c}=1$ inhibits the development of pairing 
(this value is consistent with the Thouless criterion for superconductivity~\cite{Thouless1960}). 
We remind the reader that superconductivity is suppressed by the application of a Zeeman field larger than the critical 
Clogston-Chandrasekhar value~\cite{Clogston1962}, which translates into a critical imbalance $\delta\mu_{\rm CC}=\sqrt{2}$.
Note also that the transition (\ref{eq:tzeroth4}) from the paired to the unpaired regime is continuous with 
a singularity in the derivative, as in a phase transition of the second kind. Subleading terms in the 
asymptotic expansion in powers of $1/\omega_{\rm D}$ are presented in Appendix~\ref{appendix:app_2} and 
do not modify the key features of Eq.~(\ref{eq:tzeroth4}).

We now study Eqs.~(\ref{eq:imba5}) for small but finite $\theta$ in order to determine the value of the imbalance
$\delta\mu_{\rm r}$ above which the dependence of $\gamma$ on $\theta$ ceases to be monotonic, {\it i.e.} 
\begin{eqnarray}\label{eq:mur}
\gamma(\theta, \delta\mu) < \gamma(0, \delta\mu) & \mbox{ if } & \delta\mu < \delta\mu_{\rm r} \nonumber \\
\gamma(\theta, \delta\mu) > \gamma(0, \delta\mu) & \mbox{ if}  & \delta\mu > \delta\mu_{\rm r}~.
\end{eqnarray}
We expand $\gamma$ near $\theta=0$, $\gamma(\theta, \delta\mu) = \gamma(\theta=0, \delta\mu)+\theta \gamma_{1}(\delta\mu) 
+ \theta^{2}\gamma_{2}(\delta\mu) + {\cal O}(\theta^{3})$. A similar expansion is written for $\omega(\theta, \delta\mu)$.
The integrals involving the Fermi functions in Eqs.~(\ref{eq:imba5}) can easily be computed up to 
second order in $\theta$ using the Sommerfeld method, as briefly outlined in Appendix~\ref{appendix:app_3}. 
In the limit $\omega_{\rm D}\gg 1$ we obtain $\gamma_{1}(\delta\mu)=0$ and
\begin{equation}\label{eq:gamma2}
\gamma_{2}(\delta\mu)=-\frac{2\pi^{2}}{3}\frac{1-2\delta\mu^{2}}{\sqrt{1-\delta\mu^{2}}}~.
\end{equation}
We see that $\gamma_{2}>0$ for $\delta\mu > \sqrt{2}/2$, {\it i.e.} $\delta\mu_{\rm r}=\sqrt{2}/2$ and $\gamma$ increases quadratically with temperature. In Appendix~\ref{appendix:app_3} we report an expression for $\delta\mu_{\rm r}$ that is correct up to second order 
in $1/\omega_{\rm D}$.

Before concluding this section, we would like to mention that the existence of a re-entrance for $\delta\mu>1$, 
{\it i.e.} $\partial^2 \delta\mu/\partial \theta^2|_{\gamma=0}>0$, can be proven by arguments similar to 
those that led to Eq.~(\ref{eq:gamma2}).

\subsection{Analysis of the pairing oscillations}
\label{sect:pairing_oscill}

In this section we focus on the oscillatory dinamics of the pairing function, shown in the right panels of Fig.~\ref{fig:one}.
We follow Refs.~\onlinecite{YTA2006,YAKE2005} and~\onlinecite{YKA2005} and use the formalism of the so-called Lax vector that allows an implicit analytical solution of the BdGE.

The Lax vector ${\bm L}(w)$ is a three-dimensional vector whose components are rational polynomials 
of an auxiliary complex variable $w$ and is defined as~\cite{YKA2005}
\begin{equation}\label{eq:lax3}
{\bm L}(w) = -\frac{{\bm z}}{g}+\sum_{k}\frac{{\bm S}_{k}}{w-\varepsilon_{k}}~.
\end{equation}
Here ${\bm z}$ is the unit vector in the z-direction and ${\bm S}_{k}=(S^x_k,S^y_k,S^z_k)$ is a 
three-dimensional real vector whose components are defined by $S_{k}^{x}-iS_{k}^{y}=U_{k}^{\ast}V_{k}$ 
and $2S_{k}^{z}=|V_{k}|^{2}-|U_{k}|^{2}$. According to Ref.~\onlinecite{YTA2006} 
the asymptotic time evolution of the solutions of the BdGE can be predicted by looking at the roots of $|{\bm L}(w)|^{2}$.
In the limit $N\rightarrow\infty$ almost all the roots of $|{\bm L}(w)|^{2}$ 
cluster together on the real axis. Few isolated roots with non-zero imaginary part define the 
frequencies that appear in the oscillations of $\Delta(t)$.

The vectors $\{{\bm S}_{k}, k=1...N\}$ can be interpreted as Anderson classical pseudospins~\cite{Anderson1958}. 
Each $k$-pseudospin represents the state of a Cooper pair and the initial state 
$(U_k,V_k)$ can be formally mapped onto a pseudospin chain.
In the case of the initial state written in Eq.~(\ref{eq:initial_conditions}), it is easy to 
see that a substantial probability $|V_{k}|^{2}$ to find a Cooper pair in the doubly-occupied 
state $|2\rangle$ corresponds to a very small 
probability $|U_{k}|^{2}$ to find it in the vacuum state $|0\rangle$. To simplify the expression 
of the Lax vector in Eq.~(\ref{eq:lax3}) we introduce, however, a more stringent condition. We take 
$gN|U_{k}^{\ast}V_{k}|\ll 1$, {\it i.e.} we assume that the initial pseudospins are almost 
entirely aligned in the ${\bm z}$ direction. The Lax vector then becomes
\begin{equation}\label{eq:lax6}
{\bm L}(w) 
	\simeq {\bm z}\left(-\frac{1}{g} + \sum_{i}\frac{-2S_{i}^{z}}{2\varepsilon_{i}-2w} \right)~.
\end{equation}
For $\varepsilon_{k}<0$, the pseudospins are aligned along the $+{\bm z}$ direction and represent 
doubly occupied states, while for $0<\varepsilon_{k}<\delta\mu$ the norm of the 
pseudospins $|{\bm S}_k|$ is negligible and vanishes at zero temperature, and for $\varepsilon_{k}>\delta\mu$ 
the pseudospins are aligned along the $-{\bm z}$ direction and represent vacuum states.
The length of the $k$-th pseudospin $|{\bm S}_k|$ gives the probability that the $k$-th 
subsystem is in the condensate sector $|0\rangle\otimes|2\rangle$. So the states that contain 
unpaired electrons correspond to pseudospins with smaller length.

In our case it is easy to see 
that all the roots of $|{\bm L}(w)|^2$ in Eq.~(\ref{eq:lax6}) are doubly-degenerate and are 
given by the solutions $\zeta$ of the consistency equation 
(\ref{eq:linz20}) and their complex-conjugates. 
At this point we remind the reader that in Sect.~\ref{sect:linear_instab} 
we have found a single solution 
of Eq.~(\ref{eq:linz20}) (illustrated in the top panel of Fig.~\ref{fig:three}) with non zero 
imaginary part. This implies that the root diagram of $|{\bm L}(w)|^{2}$ in the complex plane 
contains two degenerate vertical cuts.

The corresponding solution of the BdGE has the form~\cite{YAKE2005}
\begin{equation}\label{eq:numint1}
\Delta(t) = \Delta_{+} {\rm dn}((\Delta_{+}(t-t_{0}),k)~,
\end{equation}
with $k^{2} \equiv 1-\Delta_{-}^{2}/\Delta_{+}^{2}$. 
Here ${\rm dn}(x,k)$ is a Jacobi elliptic function 
and the maximum amplitude of the oscillations $\Delta_{+}$ is equal 
to the imaginary part of the root of $|{\bm L}(w)|^2$, which we have just shown to be equal to 
$\gamma=\Im m~\zeta$.
The period of the nonlinear oscillations can be written in terms  of the complete elliptic integral 
of the first kind $K(x)$ as
\begin{equation}\label{eq:period}
T=\frac{2}{\Delta_{+}}K(\sqrt{1-\Delta_{-}^{2}/\Delta_{+}^{2}})~.
\end{equation}
The parameter $\Delta_{-}$ is not fixed by this analysis and depends on the values of $S_{k}^{-}$.
The distribution of $S_{k}^{-}$ depends on the particular realization of the random phases $\phi_{k}$, 
so that we expect fluctuations in the value of $\Delta_{-}$ and $T$.

\begin{figure}
\begin{center}
\includegraphics[width=1.00\linewidth]{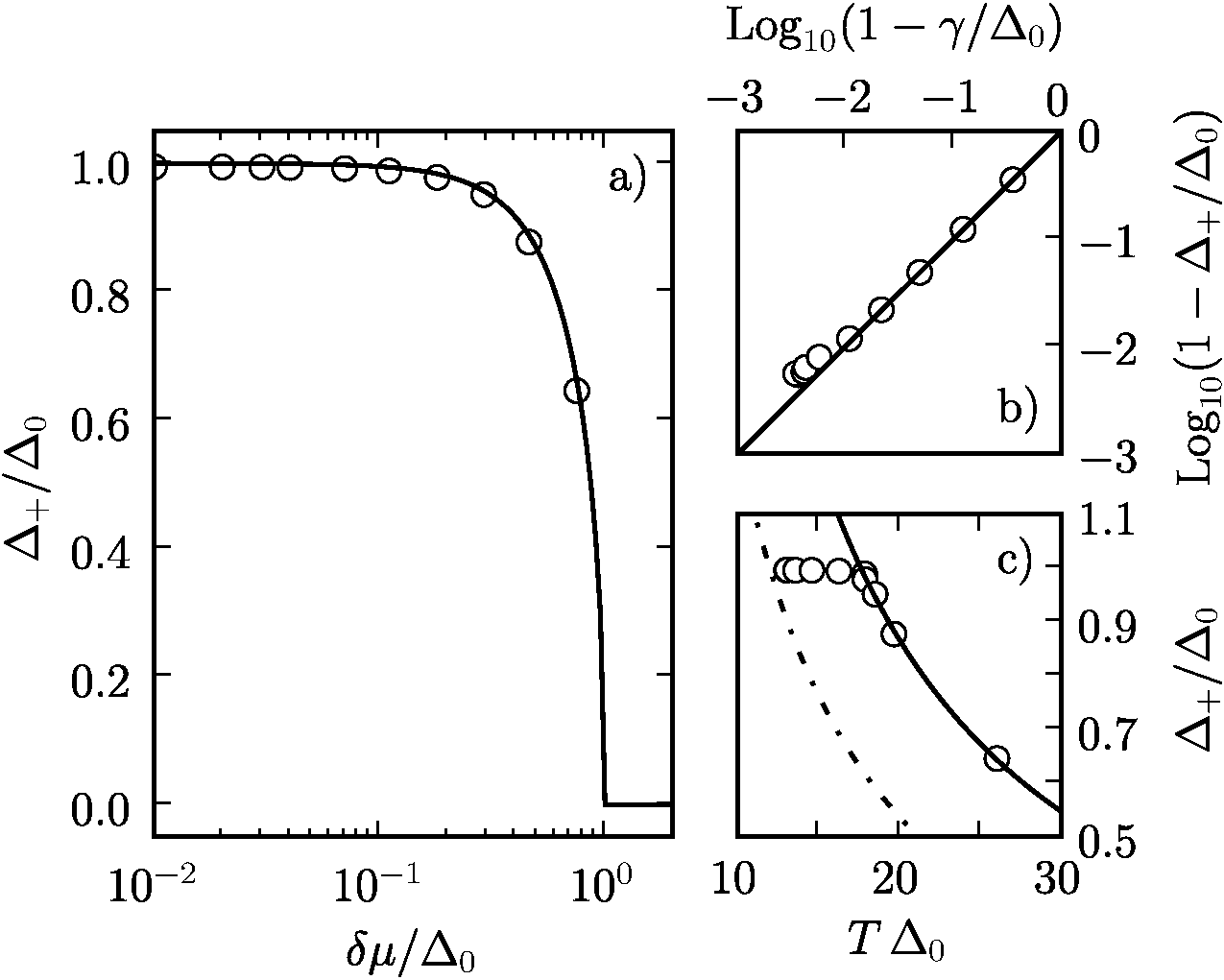}
\caption{A comparison between the results of the simulations described in Sect.~\ref{sect:numerical_simulations} (circles) and the analytical results of Sect.~\ref{sect:linear_instab} (lines). All the numerical results shown are average values over fifty simulation runs, for $\theta=10^{-2}$ (see Appendix~\ref{appendix:app_1}, Fig.~\ref{fig:three_app}).
Panel (a): the average maxima of $\Delta_+$ (circles) and the theoretical prediction given in Eq.~(\ref{eq:tzeroth4}) (solid line). Panel (b): the imaginary part $\gamma$ of the instability exponent (calculated as explained in Fig.~\ref{fig:one}) is shown to coincide with $\Delta_{+}$, the solid line being 
the $\Delta_+=\gamma$ bisector. Panel (c): the amplitude $\Delta_+$ of the oscillations (circles) is shown as a function of the period $T$. The dashed (solid) line is the period $T$ for $\Delta_-=10^{-2}$ ($\Delta_-=6\times 10^{-4}$), as from Eq.~(\ref{eq:period}).\label{fig:four}}
\end{center}
\end{figure}

In Fig.~\ref{fig:four} we show that the numerical solutions of the BdGE illustrated in 
Fig.~\ref{fig:one} agree very well both with the linear-instability analysis 
(Sect.~\ref{sect:linear_instab}) and with the analysis based on the Lax polynomial.

\section{Conclusions}
\label{sect:conclusions}

The presence of a population imbalance modifies dramatically the dynamical pairing instability in a 
two-component ultracold Fermi gas when an atom-atom attraction is suddenly switched on. 
In this work we have considered the case when the instability occurs {\it via} the s-wave pairing 
channel. We find that the dynamical instability is suppressed if the initial imbalance exceeds 
a critical temperature-dependent value, in analogy with what happens in the equilibrium situation. 
The exponent characterizing the linear-instability regime does not depend monotonically on 
temperature and shows an interesting re-entrant behavior in the temperature-imbalance plane. 
A similar behavior has been observed in equilibrium calculations since the early work of Sarma~\cite{Sarma1963}, 
though in that case the re-entrant behavior corresponds to a metastable state. 
In the dynamical situation the variational principle on the grand-canonical thermodynamic potential is of course 
not present and such re-entrant behavior can indeed be observed. It is very interesting to understand how 
our findings show up in a radio-frequency spectroscopy measurement~\cite{dzero07}.
Another important aspect, which is currently under investigation, is to understand whether it is 
possible to access more exotic pairing states after a quench.

\acknowledgments
This work was partially supported 
by a research grant of SNS and by MIUR. We wish to thank 
Pasquale Calabrese and Michael K\"{o}hl for useful discussions.
The computations have been performed with the Open Source scipy/numpy/matplotlib 
packages of the Python programming language.

\appendix

\section{Qualitative analysis of the nonlinear oscillations}
\label{appendix:app_1}

\begin{figure*}
\begin{center}
\includegraphics[width=1.00\linewidth]{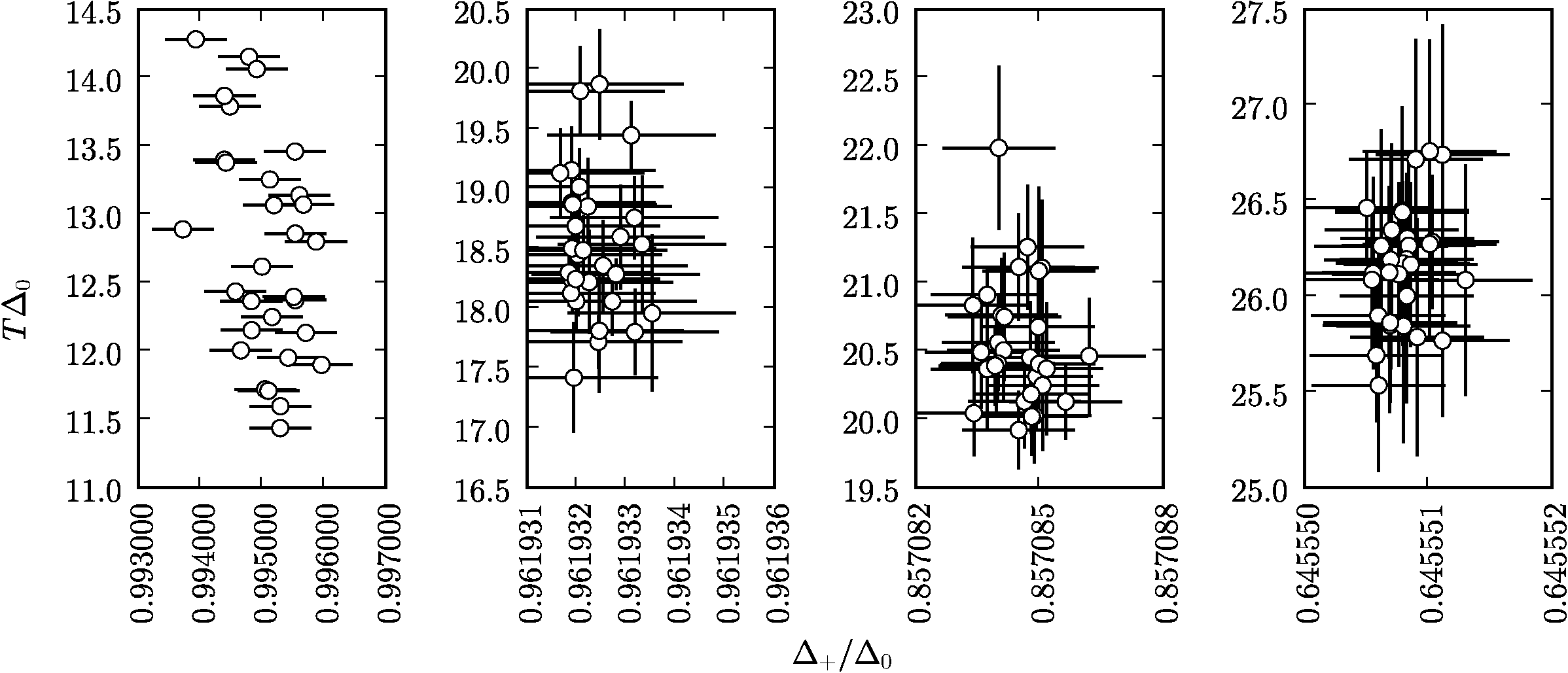}
\caption{Qualitative analysis of fluctuations in nonlinear oscillations similar to those shown in Fig.~\ref{fig:one}. 
Imbalance increases from left to right ($\delta\mu=0.0, 0.25, 0.5,$ and $0.75$). 
The coordinates of each circle give the average value of $\Delta_+$ and $T$ for one realization 
of the random initial phases $\phi_k$ with ${\cal N}_{\rm max}=10$. 
The error bars represent the standard deviations $\delta\Delta_{+}$ and $\delta T$.
\label{fig:two_app}}
\end{center}
\end{figure*}

\begin{figure}
\centering
\includegraphics[width=1.00\linewidth]{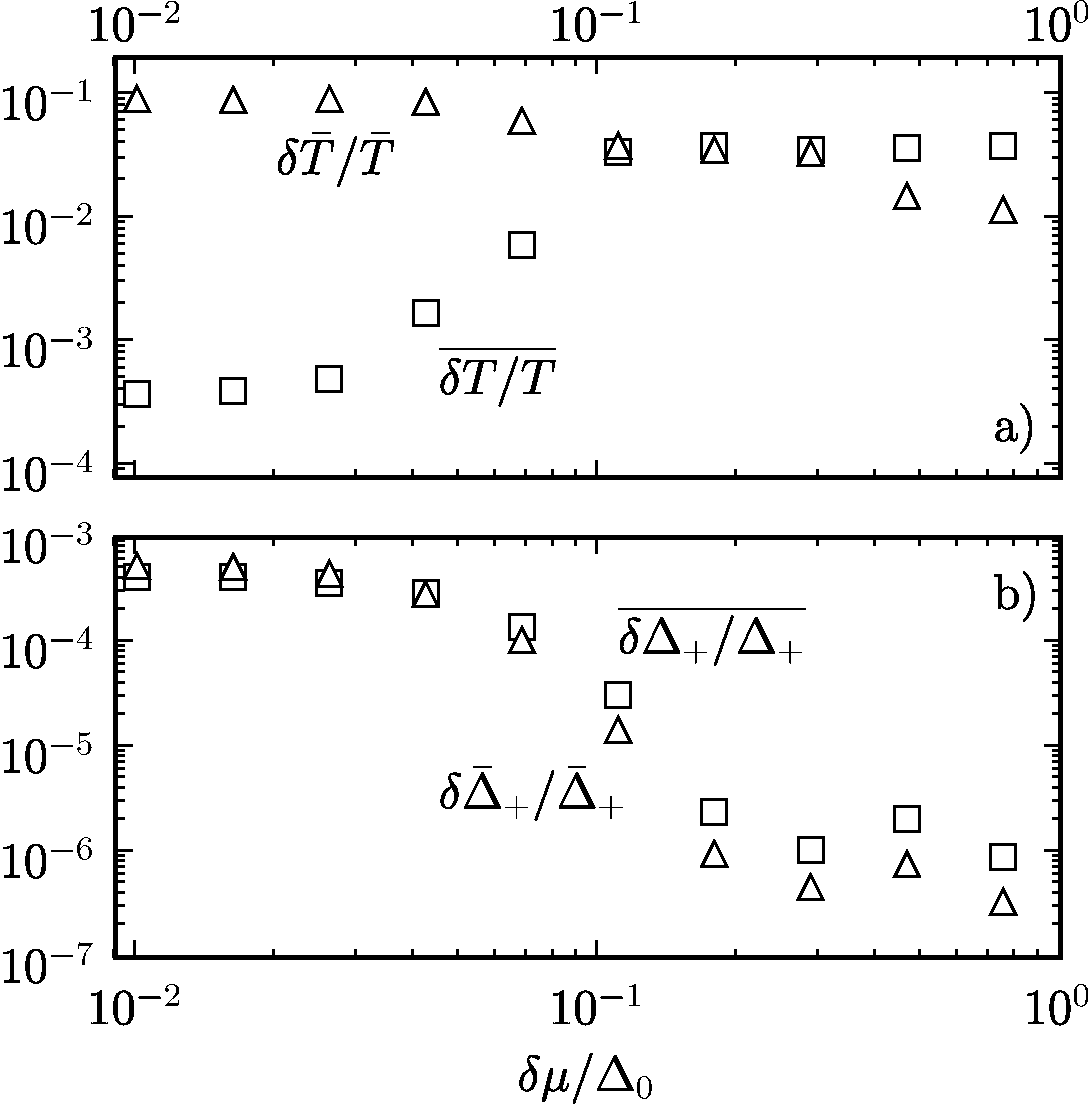}
\caption{Top panel: $\delta {\bar T}/{\bar T}$ (triangles) and $\overline{\delta T/T}$ (squares) are shown as functions of the imbalance $\delta\mu$. Bottom panel: $\delta {\bar \Delta}_+/{\bar \Delta}_+$ (triangles) and $\overline{\delta \Delta_+/\Delta_+}$ (squares) are shown as functions of the imbalance $\delta\mu$.\label{fig:three_app}}
\end{figure}

The very regular shape of the nonlinear oscillations allows us to define 
an average period $T$ and a maximum amplitude $\Delta_+$ for each simulated profile $|\Delta(t)|$. In practice, these quantities are calculated as follows. For each single realization of the random phases $\phi_k$ we find the coordinates
$\{t_{i}, \Delta_{+i}\}_{i=1}^{{\cal N}_{\rm max}}$ of the first ${\cal N}_{\rm max}$ peaks by means of a cubic interpolation. Then we compute the averages $T=\sum_i (t_{i+1}-t_{i})/({\cal N}_{\rm max}-1)$ and $\Delta_{+}=\sum_i\Delta_{+i}/{\cal N}_{\rm max}$, 
and their standard deviations $\delta T$ and $\delta \Delta_+$.

In order to illustrate the robustness of the nonlinear oscillations shown in Fig.~\ref{fig:one}, we report in Fig.~\ref{fig:two_app} an 
analysis of their shapes and periods, as found for a total of thirty
 realizations of the random phases. We notice that the spread of both $T$ and $\Delta_+$ diminishes with increasing imbalance, becoming comparable to the typical $\delta T$ and $\delta \Delta_+$ that one finds in a single realization. That is, with increasing imbalance the quantities $T$ and $\Delta_+$ become less and less dependent on the initial random phases. 

In Fig.~\ref{fig:three_app} we present a more quantitative  
account of the effect of the random initial conditions on the 
magnitude of the fluctuations. 
We have computed the average ${\bar T}$ of the period
and the corresponding standard deviation $\delta {\bar T}$ over fifty realizations. 
We see that the relative fluctuations $\delta {\bar T}/{\bar T}$ 
drop by one order of magnitude when $\delta\mu$ spans the range $[10^{-2},1]$ (see the top panel in Fig.~\ref{fig:three_app}). The average $\overline{\delta T/T}$ of the relative fluctuations of the period increases instead by two orders of magnitude when $\delta\mu$ spans the range $[10^{-2},10^{-1}]$, while it becomes comparable to $\delta {\bar T}/{\bar T}$ for $\delta\mu \gtrsim 10^{-1}$.

Finally, in the bottom panel of Fig.~\ref{fig:three_app} we illustrate the behavior of the relative fluctuations $\delta {\bar \Delta}_+/{\bar \Delta}_+$ of the amplitudes, which drop by three orders of magnitude when $\delta\mu$ spans the range $[10^{-2},1]$. The average $\overline{\delta \Delta_+/\Delta_+}$ of the relative fluctuations of the amplitude remains of the same order of magnitude as $\delta {\bar \Delta}_+/{\bar \Delta}_+$.

\section{Critical imbalance at zero temperature}
\label{appendix:app_2}

In this Appendix we determine analytically the critical imbalance $\delta\mu_{\rm c}$ at zero temperature, defined by $\gamma(\theta=0, \delta\mu_{\rm c}) = 0$. The value $\omega_{\rm c}$ of $\omega$ at criticality has also to be determined to solve consistently Eq.~(\ref{eq:imba5}).

Equation (\ref{eq:imba5}) at $\theta=0$ reads
\begin{subequations}\label{eq:imba5_0}
\begin{eqnarray}
& [(2\omega_{\rm D}-\omega)^{2}+\gamma^{2}]
	[(2\omega_{\rm D}+\omega)^{2}+\gamma^{2}] = & \nonumber \\
& = (\omega_{\rm D} + \sqrt{1+\omega_{\rm D}^{2}})^{4}
	[(2\delta\mu-\omega)^{2}+\gamma^{2}]
	[\omega^{2}+\gamma^{2}] &\nonumber\\ 
	\label{eq:tzero8a}
\end{eqnarray}
and
\begin{eqnarray}
& \arctan[(2\omega_{\rm D}-\omega) / \gamma] -
	\arctan[(2\delta\mu-\omega ) / \gamma] = & \nonumber \\
& = -\arctan(\omega  /\gamma)
	+\arctan[(2\omega_{\rm D}-\omega)/\gamma]~. & \label{eq:tzero8b}
\end{eqnarray}
\end{subequations}
We assume that $2\delta\mu_{\rm c}>\omega_{\rm c}$, as is suggested by the numerical solution and also 
by the zeroth-order solution (\ref{eq:tzeroth4}). Then in Eq.~(\ref{eq:tzero8a}) we put $\gamma=0$ and obtain
\begin{equation}\label{eq:tzerocr6}
2\delta\mu-\omega  = 
	\frac{4\omega_{\rm D}^{2}-\omega^{2}}
	{\omega(\omega_{\rm D}+\sqrt{1+\omega_{\rm D}^{2}})^{2}}~.
\end{equation}
In Eq.~(\ref{eq:tzero8b}) we perform the limit $\gamma\to 0$ and find
\begin{equation}\label{eq:tzerocr4}
\frac{1}{\omega} - \frac{1}{2\omega_{\rm D}+\omega} = 
	\frac{1}{2\delta\mu-\omega} - \frac{1}{2\omega_{\rm D}-\omega}~.
\end{equation}
In deriving this result we have used that $\arctan{(a/\gamma)} \to \pi/2 - \gamma/a$.

Substituting Eq.~(\ref{eq:tzerocr6}) into Eq.~(\ref{eq:tzerocr4}) we obtain
\begin{equation}\label{eq:tzerocr9.1}
\omega^2_{c} = \frac{2}{\displaystyle
1+\sqrt{1+1/\omega_{\rm D}^{2}}}~.
\end{equation}
Using this result back into Eq.~(\ref{eq:tzerocr6}) we find 
\begin{equation}\label{eq:tzerocr12}
\delta\mu_{\rm c} = \frac{2\omega_{\rm D}}{\omega_{\rm c}/\omega_{\rm D}+\omega_{\rm D}/\omega_{\rm c}}~.
\end{equation}
A second-order expansion of Eqs.~(\ref{eq:tzerocr9.1}) and ~(\ref{eq:tzerocr12}) 
in powers of $1/\omega_{\rm D}$ finally gives
\begin{equation}\label{eq:tzerocr10}
\omega_{\rm c} \simeq
	1 - \frac{1}{8}\frac{1}{\omega^2_{\rm D}},
	\quad
\delta\mu_{\rm c} \simeq
	1-\frac{3}{8}\frac{1}{\omega^2_{\rm D}}~.	
\end{equation}
In our computations $\omega_{\rm D}=5.0$, so these second order corrections are 
of order $10^{-2}$ ($\omega_{\rm c}\simeq 0.995$ and $\delta\mu_{\rm c}\simeq 0.985$).

Now we show that the slope of the curve $\gamma(\theta=0,\delta\mu)$ is singular at the critical imbalance $\delta\mu_{\rm c}$ and we find an asymptotic form for the profile. 
We make the {\it Ansatz} $\gamma=\alpha \sqrt{\delta\mu_{\rm c}-\delta\mu}$ and 
$\omega=\omega_{c}+\kappa (\delta\mu_{\rm c}-\delta\mu)$. 
Substituting this  into Eqs.~(\ref{eq:imba5_0}) and discarding powers of $\delta\mu_{\rm c}-\delta\mu$ higher than one we find 
that the {\it Ansatz} is consistent provided that 
\begin{equation}\label{eq:steepness}
\alpha^2 = 2\omega_{\rm c}\left ( 1+\frac{\omega_{\rm c}^{2}}{4\omega_{\rm D}^{2}} \right )\simeq 2 \left(1+\frac{1}{8}\frac{1}{\omega_{\rm D}^{2}}\right)~.
\end{equation}

\section{Subleading corrections to $\delta\mu_{\rm r}$}
\label{appendix:app_3}

In the main body of the paper, immediately above Eq.~(\ref{eq:gamma2}), we introduced an expansion of $\gamma$ in powers of temperature near $\theta=0$. The coefficient $\gamma_{1}(\delta\mu)$ of the linear term is identically zero, while the coefficient $\gamma_{2}(\delta\mu)$ of the quadratic term has been given only for $\omega_{\rm D}\to \infty$.
The equation $\gamma_2(\delta\mu_{\rm r})=0$ defines the imbalance $\delta\mu_{\rm r}$ above which the dependence of $\gamma$ on $\theta$ ceases to be monotonic. 
In this Appendix we find the second-order corrections to the quantity $\delta\mu_{\rm r}$ in powers of $1/\omega_{\rm D}$.

To this end, we note that Eq.~(\ref{eq:imba5}) can be written 
in the general form
\begin{equation}\label{eq:sommerfeld}
\int_{-\omega_{\rm D}}^{\omega_{\rm D}}d\varepsilon~
g(\varepsilon)[1-f(\varepsilon)-f(\varepsilon-\mu)] = K~,
\end{equation}
with $f(x) = 1/(e^{\beta x}+1)$. To compute $\gamma_2(\delta\mu)$ we need to expand this equation in powers of the temperature $\theta$. In order to do so we follow a familiar Sommerfeld procedure:
we perform an integration by parts in Eq.~(\ref{eq:sommerfeld}), expanding the primitive $G(\varepsilon)$ of $g(\varepsilon)$ in powers of $\theta$. 
The Sommerfeld expansion of the integrals involving the Fermi-Dirac functions to order $\theta^2$ gives
\begin{eqnarray}\label{eq:tlim7}
G(\omega_{\rm D})&+&G(-\omega_{\rm D})
	-G(0) -G(\delta\mu)\nonumber\\
	&-&\theta^{2}\frac{\pi}{3}
\left[\left.\frac{\partial^2 G(\varepsilon)}{\partial \varepsilon^2}\right|_{\varepsilon=0} 	+
	\left.\frac{\partial^2 G(\varepsilon)}{\partial \varepsilon^2}\right|_{\varepsilon=\delta\mu}\right] = K~.\nonumber\\
\end{eqnarray}
For the first of the two Eqs.~(\ref{eq:imba5}) the function $G$ is given by
\begin{equation}
G(x)=\frac{1}{4}\ln{[(2x-\omega)^{2}+\gamma^{2}]}~,
\end{equation}
while for the second it is given by
\begin{equation}
G(x) = \frac{1}{2\gamma}\arctan{\left(\frac{2x-\omega}{\gamma}\right)}~.
\end{equation}
We remark that $G$ depends parametrically on the temperature $\theta$ through the functions 
$\omega=\omega(\theta, \delta\mu_{\rm r})$ and $\gamma=\gamma(\theta, \delta\mu_{\rm r})$. 
We expand Eq.~(\ref{eq:tlim7}) order by order in powers of $\theta$ and subsequently in powers of 
$1/\omega_{\rm D}$. By imposing that $\gamma_{2}(\delta\mu_{\rm r})=0$ we finally obtain
\begin{equation}
\delta\mu_{\rm r}\simeq \frac{\sqrt{2}}{2}
\left(1+\frac{1}{4}\frac{1}{\omega_{\rm D}^{2}}\right)~.
\end{equation}


\begin{thebibliography}{77}
\bibitem{footnote1}
	In the standard BCS formulation $\omega_{\rm D}$ is the Debye frequency associated
	with the presence of phonon-mediated electron-electron interactions.
	In the present problem $\omega_{\rm D}$ is solely an ultraviolet cutoff on which the
	physical results such as the dynamical ``phase diagram'' in Fig.~\ref{fig:three} depend
	only very weakly.
\end{thebibliography}
\end{document}